\documentclass[usenatbib]{mn2e}
\usepackage{epsfig,times,lscape,aas_macros,amsmath}
\usepackage{epstopdf}

\title{The particle content of low-power radio galaxies in groups and clusters} \author[J. H. Croston \& M. J. Hardcastle]
{J. H. Croston$^{1}$\thanks{Email: J.Croston@soton.ac.uk}
and M.J. Hardcastle$^{2}$\\$^1$ School of Physics and Astronomy,
University of Southampton, Highfield, Southampton SO17 1BJ, UK\\$^2$
School of Physics, Astronomy and Mathematics, University of
Hertfordshire, College Lane, Hatfield, Hertfordshire AL10 9AB, UK}
 
\pagerange{\pageref{firstpage}--\pageref{lastpage}}
\pubyear{2009}
\begin{document}

\maketitle

\label{firstpage}

\begin{abstract}

The synchrotron-radiating particles and magnetic fields in low-power
radio galaxies (including most nearby cluster-centre sources), if at
equipartition, can provide only a small fraction of the total internal
energy density of the radio lobes or plumes, which is now well
constrained via X-ray observations of their external environments. We
consider the constraints on models for the dominant energy
contribution in low-power radio-galaxy lobes obtained from a detailed
comparison of how the internal equipartition pressure and
external pressure measured from X-ray observations evolve with
distance for two radio galaxies, 3C\,31 and Hydra A. We rule out
relativistic-lepton dominance of the radio lobes, and conclude that
models in which magnetic field or relativistic protons/ions carried up the
jet dominate lobe energetics are unlikely. Finally, we argue that
entrainment of material from the jet surroundings can provide the
necessary pressure, and construct a simple self-consistent model of
the evolution of the entrainment rate required for pressure balance
along the 100-kpc scale plumes of 3C\,31. Such a model requires that
the entrained material is heated to temperatures substantially above
that of the surrounding intra-group medium, and that the temperature
of the thermal component of the jet increases with distance, though
remaining sub-relativistic.

\end{abstract}

\begin{keywords}
galaxies: active -- X-rays: galaxies: clusters
\end{keywords}

\section{Introduction}
\label{intro}
Low-power (FRI: \citealt{fanaroff74}) radio galaxies are commonly
found in the centres of rich galaxy groups and clusters, where they
are thought to play an important role in regulating the central gas
properties and galaxy evolution via a (currently poorly understood)
feedback process \citep[e.g.][ and references
  therein]{mcnamara07,fabian12}. Among the many uncertainties about the way in
which this feedback process operates, one long-standing problem is the
unknown nature of the dominant particle or field component within the
radio lobes, which are important as the lobes are the means of energy
transfer to the surrounding gas via their expansion. The radio
synchrotron emission from the lobes provides only a combined
constraint on electron density and magnetic field strength, and so it
has been common to assume equipartition of energy in field and
radiating particles \citep[e.g.][]{burbidge56}, which corresponds
roughly to the minimum total energy the source requires in order to
produce the observed radio emission. But while the lobes of powerful FRII
radio galaxies appear to be close to equipartition
\citep[e.g.][]{croston05a,kataoka05}, it has been known for some time
that the energy content of FRI radio galaxies must be distributed
differently to that of FRIIs, as the radiating particles and magnetic
field, if at equipartition, cannot in the vast majority of cases
provide sufficient pressure to balance the measured external pressures
surrounding FRI lobes \citep[e.g.][]{morganti88,worrall00}.

The external pressure acting on the jets and lobes can now be
constrained tightly on scales of a few to several hundred kpc for many
low-power radio galaxies, using X-ray observations of the surrounding
group or cluster gas with {\it Chandra} and {\it XMM-Newton}
\citep[e.g.][]{hardcastle02b,croston03b,croston08a}. If it is assumed that the jets and lobes
are close to pressure equilibrium with the surrounding medium (likely to be
true on kpc -- hundred kpc scales for low-power sources), then the
external pressure profile must correspond closely to the run of
internal pressure along the jet as it evolves into a lobe or plume.
The internal pressure cannot be measured directly from the radio
observations of the source; however, the internal pressure in some
combination of radiating particles (electrons and positrons) and
magnetic field can be measured by modelling the radio emission. This
type of comparison has now been carried out for many low-power radio
galaxies, including large samples of cavity sources in galaxy clusters
(including so-called ``ghost'' cavities in which any radio emission is
weak or absent), and, as mentioned above, typically shows that the
radiating particles and magnetic field cannot dominate the internal
energy of the source {\it if they are at equipartition}
\citep{croston03b,croston08a,dunn04,dunn05,dunn06b,birzan08}.

Given that the lobes of low-power radio galaxies cannot be dominated
by an equipartition electron-positron plasma, other models for the
energetically dominant component of the radio lobe contents must be
considered. The two most obvious explanations are that the dominant
internal pressure is provided by a departure from equipartition or by
a significant population of non-radiating particles. There is evidence
from X-ray inverse Compton observations that powerful FRII radio
galaxies may deviate from equipartition by a small amount in the
direction of electron dominance \citep[e.g.][]{isobe02,croston05a}; however,
electron dominance by large factors would be expected to produce
detectable levels of X-ray inverse-Compton emission in at least some
FRI radio galaxies, which are inconsistent with observations
\citep{hardcastle98e,croston03b}. Recently, detailed models of magnetically dominated
jets and lobes have been developed \citep[e.g.][]{li06,nakamura06};
however, they are difficult to reconcile with observations, e.g. of
radio jet polarization properties and geometry (see later discussion).
Proton-dominated models have been discussed by a number of authors
\citep[e.g.][]{deyoung06,birzan08,mcnamara07}, but it is energetically difficult
to supply the proton population required by transport from the inner
jet \citep[e.g.][]{deyoung06}.

There are several reasons to favour instead a model in which
entrainment of material as the jet expands leads to an
energetically dominant proton population on scales of tens to hundreds
of kpc. Entrainment of the ISM and ICM is thought to be the means by
which FRI jets decelerate from relativistic to transonic speeds on kpc
scales \citep[e.g.][]{bicknell94}. There is growing observational evidence
that entrainment is occurring \citep[e.g.][]{hardcastle03b,hardcastle07a}, as well as
strong support for its importance from detailed kinematic modelling of
FRI jets \citep[e.g.][]{laing02b,laing06a}. A model in which entrainment
accounts for the apparent ``missing'' pressure in FRI radio galaxies
also has the advantage of explaining the observed difference in the
energetics of the FRI and FRII populations (the former being massively
underpressured if at equipartition, while the latter appear close to
equipartition both from IC observations and pressure comparisons)
without the need to invoke differences in the intrinsic particle
content of the inner jets, which might require different jet production
mechanisms: since FRII jets do not decelerate or interact with their
environments significantly, they would not in general be expected to
entrain significant amounts of material. Finally, we have previously
found a relationship between FRI source structure and particle/energy
content, suggesting that sources likely to be undergoing strong
entrainment have a larger contribution from non-radiating material
than those likely to be weakly entraining \citep{croston08a}. This provides
further support for an entrainment-dominated model.

In this paper we investigate in detail the observational constraints
on models for the particle and energy content of low-power radio
galaxies, by considering how the non-radiating and radiating
components of the jets in the well-studied radio galaxy 3C\,31
must evolve with distance in order to maintain pressure balance and
produce the observed radio emission. We use new deep X-ray data and
high-resolution radio data to place tight constraints on the external
pressure and internal pressure from radiating particles and field
within the radio jets and plumes of 3C\,31 . We consider in detail the constraints this
result provides for what particle population or magnetic field
structure dominates the source energetics, and
also carry out a pressure comparison for the cluster-centre source
Hydra\,A  as a preliminary test of the generality of our results.

Throughout the paper we use a cosmology in which $H_0 = 70$ km
s$^{-1}$ Mpc$^{-1}$, $\Omega_{\rm m} = 0.3$ and $\Omega_\Lambda =
0.7$. At the redshifts of 3C\,31 ($z = 0.0169$) and Hydra\,A ($z = 0.0549$),
this gives luminosity distances of $D_{L} = 73.3$ Mpc and $D_{L} = 244.9$
Mpc, respectively, and angular scales of $0.3438$ kpc/arcsec (3C\,31)
and $1.067$ kpc/arcsec (Hydra\,A). Spectral indices $\alpha$ are defined in
the sense $S_{\nu} \propto \nu^{-\alpha}$. Reported errors are
1$\sigma$ for one interesting parameter, except where otherwise noted.

\section{Observational constraints}
\label{obs}
\subsection{External pressure of the hot-gas environment}

\subsubsection{3C\,31}

We used new {\it XMM-Newton} data to obtain a radial profile of the
external pressure surrounding the radio jets and plumes in 3C\,31. We
observed 3C\,31 on 2008 July 1st for $\sim 50$ ks (ObsID
0551720101). The data were processed in the standard way using {\it
XMM-Newton} SAS version 11.0.0, and the latest calibration files from
the {\it XMM-Newton} website. The pn data were filtered to include
only single and double events (PATTERN $\leq 4$), and FLAG==0, and the
MOS data were filted according to the standard flag and pattern masks
(PATTERN $\leq 12$ and \#XMMEA\_EM, excluding bad columns and
rows). Unfortunately the observation was badly affected by background
flares, and so after filtering for good time intervals, the remaining
clean exposure durations were 24, 29, and 24 ks for the MOS1, MOS2 and
pn cameras, respectively.

 Surface brightness profiles in the energy range 0.3 -- 5.0 keV were
 extracted from the {\it XMM-Newton} data using the closed-filter
 double-background method described by \citet{croston08a}. The {\it
   Chandra} surface brightness profile of \cite{hardcastle02b} was also used to
 help constrain the inner profile shape. The combined {\it XMM-Newton}
 (MOS1, MOS2 and pn) profile and {\it Chandra} profile were jointly
 fitted with a projected double beta model \citep{croston08a}, convolved
 with the appropriate point-spread function (PSF) for each telescope,
 using the Markov-Chain Monte Carlo (MCMC) method described by
 \citet{ineson13}. The resulting model was used to obtain a gas density
 profile for the environment.

A corresponding temperature profile was obtained by extracting spectra
from six annular regions, and using the background fitting method
described by \citet{croston08a}, which correctly accounts for both
particle and X-ray background, to obtain (projected) temperature
measurements. For each region, the spectra from the three {\it
  XMM-Newton} cameras were fitted jointly with an {\it apec} model
(using the energy range 0.3 -- 7.0 keV, but excluding the region
between 1.4 -- 1.6 keV, which is affected by an instrumental
line). The normalizations for the three cameras were allowed to vary,
but the temperatures were tied together. A free abundance fit led to
unphysically large values for the abundance, and so we fixed the
abundance to the best-fitting abundance from a global spectral fit ($Z
= 0.3$). The results of spectral fitting are given in
Table~\ref{spectra}. For the inner regions of the group, we used the
{\it Chandra} temperature profile of \citet{hardcastle02b} in order to
obtain more accurate pressure constraints. We used the deprojected
temperature values, although the effect of deprojection on the
temperature profile is small. In the outer regions of the group the
temperature varies only by $\sim 20$ per cent, so that any uncertainty
from not correcting for projection is small, and less than the
statistical uncertainty on the outer temperature.

In order to obtain a gas pressure profile with high resolution, we
fitted the measured temperature profile with the analytic model of
Vikhlinin et al. (2006), and obtained a finely binned look-up table
for $\Lambda (T)$, the conversion factor between volume emission
measure and gas density (obtained from {\sc xspec}). The resulting
table was used together with the analytic temperature model to obtain
a gas pressure profile, which is shown in Fig~\ref{press1}.

\subsubsection{Hydra\,A}

Although the majority of this paper focuses on 3C\,31, using our new
X-ray data, we also carried out a pressure comparison for Hydra\,A as
a preliminary test of whether our findings are likely to apply widely
to FRI radio galaxies. For Hydra A, we did not reanalyse the archival
{\it Chandra} and {\it XMM-Newton} observations, but made use of
previously published gas density and temperature profiles. For the gas
density profile we used the double beta model of \citet{wise07},
normalised to the density profile published by \citet{david01}. We
interpolated over the (projected) temperature profile of \citet{david01}
to obtain a cluster pressure profile over the radial ranges of
interest, which is also shown in Fig.~\ref{press1}.

\begin{table}
\caption{Results of spectral fitting for the environment of
  3C\,31. Spectral fits were obtained for annular regions between the
  radii listed, using an {\sc apec} model, were in the energy range
  $0.3 - 7$ keV, assuming $N_{H} = 5.4 \times 10^{20}$ cm$^{-2}$. The abundance was fixed to the best-fitting value from a global spectral fit, as free abundance fits led to unphysical values (likely due to the additional free parameters of the background model).}
\label{spectra}
\begin{tabular}{lrrr}
\hline
Region&$kT$&$Z$&$\chi^{2}$ (d.o.f.)\\
\hline
60 -- 80 arcsec&$1.58\pm0.1$&0.1&64 (65)\\
80 -- 120 arcsec&$1.62^{+0.6}_{-0.7}$&0.3&157 (167)\\
120 -- 200 arcsec&$1.60^{+0.5}_{-0.6}$&0.3&466 (395)\\
200 -- 300 arcsec&$1.54\pm0.5$&0.3&722 (608)\\
300 - 450 arcsec&$1.36^{+0.2}_{-0.1}$&0.3&1046 (835)\\
450 -- 600 arcsec&$1.09^{+0.12}_{-0.01}$&0.3&1191 (974)\\
\hline
\end{tabular}
\end{table}

\subsection{Internal pressure from radiating particles and magnetic field}
\label{radio}
The radio emission from the sources does not place a constraint on the
total internal pressure of the lobes, as the radiating plasma could be
far from equipartition; however, it does place constraints on the
internal pressure of the radiating particles and magnetic field. We
used high-resolution radio data to obtain profiles of synchrotron
emissivity along the jets.

For 3C\,31 we used the combined 1.4-GHz map of \citet{laing08a}, which
has a resolution of 5 arcsec, and a 330-MHz map made in the standard
way from VLA archival data in the B and C configurations (Program
AL597), with a resolution of 21.3 arcsec $\times$ 18.2 arcsec, to
obtain the most reliable low-frequency measurements for outer
regions. These data enable us to measure the source geometry and radio
surface brightness accurately in the inner regions, while adequately
sampling the source structure out to the hundred kpc scale regions of
interest. For Hydra A we used the 330-MHz map of \citet{laing04} for
the outer lobes with a resolution of 15 arcsec, and to image the inner
structure with sufficient resolution for our geometric measurements we
made a map using A and B configuration archival VLA data at 1.4 GHz
\citep[e.g.][]{taylor90} with resolution of 1.4 arcsec. For 3C\,31,
roughly 20 regions per jet were used to measure the radio flux density
and jet geometry. For Hydra\,A around 20 regions were used to study
the northern jet. In the case of 3C\,31, where significant jet bending
occurs on the scales of interest, we measured the distances along the
projected jet paths as the best way of estimating the distance
travelled by material at a particular position along the jet; however,
for simplicity we assumed initially the source is in the plane of the
sky, and does not change position angle relative to the plane. The
external pressure acting at a particular position is therefore assumed
to be the pressure at the distance corresponding to the projected
radial distance from the AGN nucleus and group centre. The effects of
projection and jet bending on our results are discussed in
Section~\ref{geometry}.

To investigate the energetics of the radiating particles and magnetic
field, we initially assumed a single electron energy distribution
consisting of a power law with spectral index of 0.55, minimum
electron energy of $\gamma=10$ and maximum energy of
$\gamma=10^{4}$. This correctly describes the radio spectra of the two
sources in the inner regions (in the GHz radio regime), but is a
somewhat flatter spectral index than is measured in the outer parts of
the source. Any systematic effects of spectral steepening at GHz
frequencies on our pressure results for the outer parts of the sources
will be small, as the total electron energy content is dominated by
the low-energy electron population. Allowing the spectral index to
vary based on the observed spectral index at GHz frequencies would
introduce large systematic uncertainty in the low-energy electron
density. We therefore used a single electron distribution for all
regions, normalized to the measured radio flux density for that region
from the appropriate radio map (in the case of Hydra A flux densities
at 5 GHz were used in order to have sufficient spatial resolution out
to a distance of 40 kpc, with the 330-MHz map used beyond that
distance). In future work we will make use of new low-frequency data
from the Low-Frequency Array (LOFAR) to improve our spectral model.

In order to investigate the variations in internal conditions along
the source, a power law was fitted to the emissivity distribution so
as to provide a smooth model for the variation with distance. Although
there is some small systematic deviation of the observed emissivity
about the model, the measured profile is never more than $\sim 40$
percent different from the model (and typically within 10
percent). 

Using the smoothed emissivity profiles for 3C\,31, we first determined
the internal pressure as a function of position along the jet, under
the assumptions of equipartition of energy between particles and
magnetic field, and no non-radiating particles ($\kappa=0$, where
$\kappa = U_{NR}/U_{R}$, i.e.  the ratio of energy density in
non-radiating particles to that in synchrotron-emitting
particles). The internal, equipartition pressure profiles for are
shown in Fig.~\ref{press1}, together with the external pressure
profiles determined from the X-ray observations. For Hydra\,A, we
simply calculated an internal pressure profile for the existing radio
bins, under the same assumptions. This profile is shown in
Fig.~\ref{press1}, illustrating a strong qualitative similarity to the
behaviour of the 3C\,31 jets.

\section{Implications for lobe contents}
\label{imp}
\begin{figure*}
\centerline{\hbox{
  \includegraphics[height=.35\textheight]{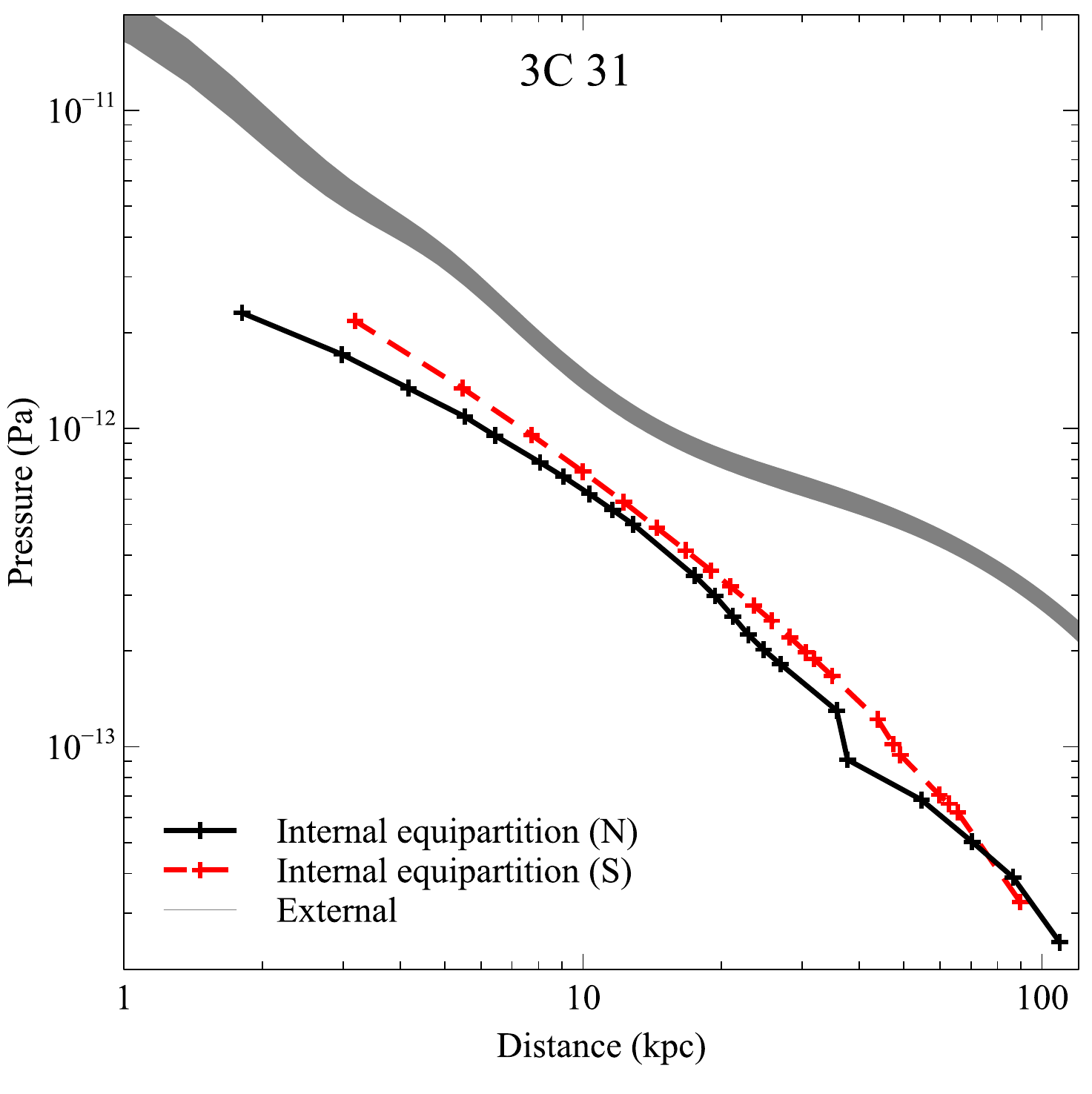}
\hskip 1.0cm
  \includegraphics[height=.35\textheight]{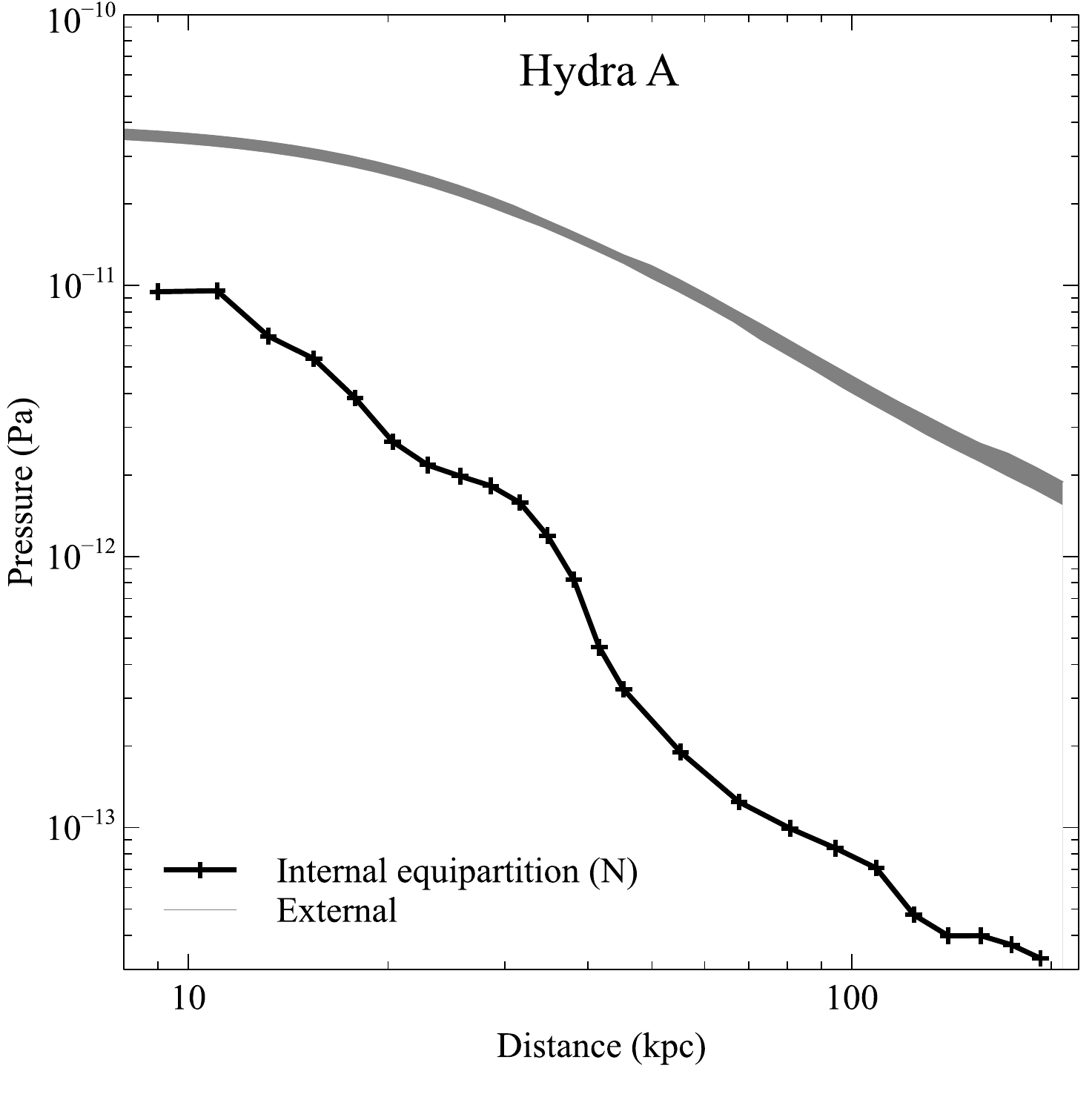}}}
  \caption{External and internal (equipartition) pressure profiles for
    the two sources 3C\,31 (l) and Hydra A (r). The external pressures
    derived from X-ray measurements are shown as shaded regions, which
    indicate the 1$\sigma$ errors, and the internal, equipartition
    pressures with the assumption of no protons are given by the solid
    black (3C31\,north, Hydra\,A north) and dashed red (3C\,31 south)
    lines. The statistical uncertainties on the internal pressures are
    negligible compared to model assumptions and so are not plotted.}
  \label{press1}
\end{figure*}

Fig.~\ref{press} shows the ratio of external pressure to internal,
equipartition pressure (with no protons) for 3C\,31, determined from the external and internal pressure
profiles described in the previous section. As seen in previous work
\citep[e.g.][]{worrall00}, the internal equipartition pressure is
significantly below the external pressure at all radii. It also is
readily apparent that the apparent pressure ``deficit'' increases with
distance, apart from in the inner $\sim 10$ kpc. This
figure illustrates clearly that on scales $> 10$ kpc
the contribution of the radiating material to the total internal
pressure of the radio source, in the equipartition case, must decrease
substantially as the jet evolves out into the group or cluster
environment. Alternatively, if equipartition between radiating
particles and magnetic field does not hold, then there must be a
systematic departure from this condition that increases with distance
from the nucleus and group/cluster centre. Such an effect was first
observed in ROSAT environmental studies
\citep[e.g.][]{hardcastle98e,worrall00}, and is also seen in our
combined {\it Chandra} and {\it XMM-Newton} analysis of NGC\,6251
\citep{evans05} and 3C\,465 \citep{hardcastle05c}; however, the higher
quality of the X-ray and radio pressure constraints in the new work we
present here places the result on a much firmer footing.

We have considered in detail the possible effects of projection on
this conclusion (see Section~\ref{geometry}). Neither 3C\,31 or Hydra\,A is
thoughout to be highly projected, and for plausible jet orientations
the plots in Fig.~\ref{press1} and ~\ref{press} do not alter
significantly as the two effects of projection act in the same
direction: the internal pressure decreases with $\theta_{los}$ since
the jet volume at a given projected distance increases, and the
external pressure acting on the jet at this projected distance
decreases because it is further out in the X-ray atmosphere whose
pressure is dropping off.

Although such detailed pressure profile comparisons have not been
carried out previously, it is interesting to note that a similar
behaviour can be seen at a statistical level in the sample of cluster
cavities studied by \citet{dunn05}, where the so-called ``ghost''
cavities are typically at much larger distances from the cluster
centre than the active lobes, which are systematically closer to
pressure balance assuming $\kappa=0$.

The pressure constraints shown in Fig.~\ref{press} can be used to test
a range of models that have been proposed for the particle or field
content dominating the energy budget of low-power radio lobes. In the
following section we consider four models for the dominant energy
content of the lobes:
\begin{itemize}
\item {\bf Model I -- lepton dominance:} the jets and lobes are out of equipartition, but the contribution from protons remains negligible and it is the radiating electrons and positrons that dominate the internal pressure
\item {\bf Model II -- magnetic field dominance:} the jets and lobes are out of equipartition, but the contribution from protons remains negligible and the magnetic field dominates the internal pressire.
\item {\bf Model III -- relativistic proton or ion dominance:} the
  jets and lobes are in equipartition, with relativistic protons (and/or ions) dominating the internal pressure (i.e. $\kappa >> 0$)
\item {\bf Model IV -- thermal gas dominance (entrainment):} the jets
  and lobes are in equipartition, with thermal material, likely
  entrained from the surrounding intragroup medium, dominating the
  internal pressure (i.e. $\kappa >> 0$)
\end{itemize}
It is clear that more complex models are possible -- in particular, it
is plausible that non-radiating particles are present, but the jets and lobes are
  not at equipartition in all locations along the jet. Such models
are harder to test, and so we begin by considering the four simpler
models listed above.

\subsection{Departures from equipartition (Models I and II)}
\label{sec:depeq}
\begin{figure}
% \centerline{\hbox{
  \includegraphics[height=.3\textheight]{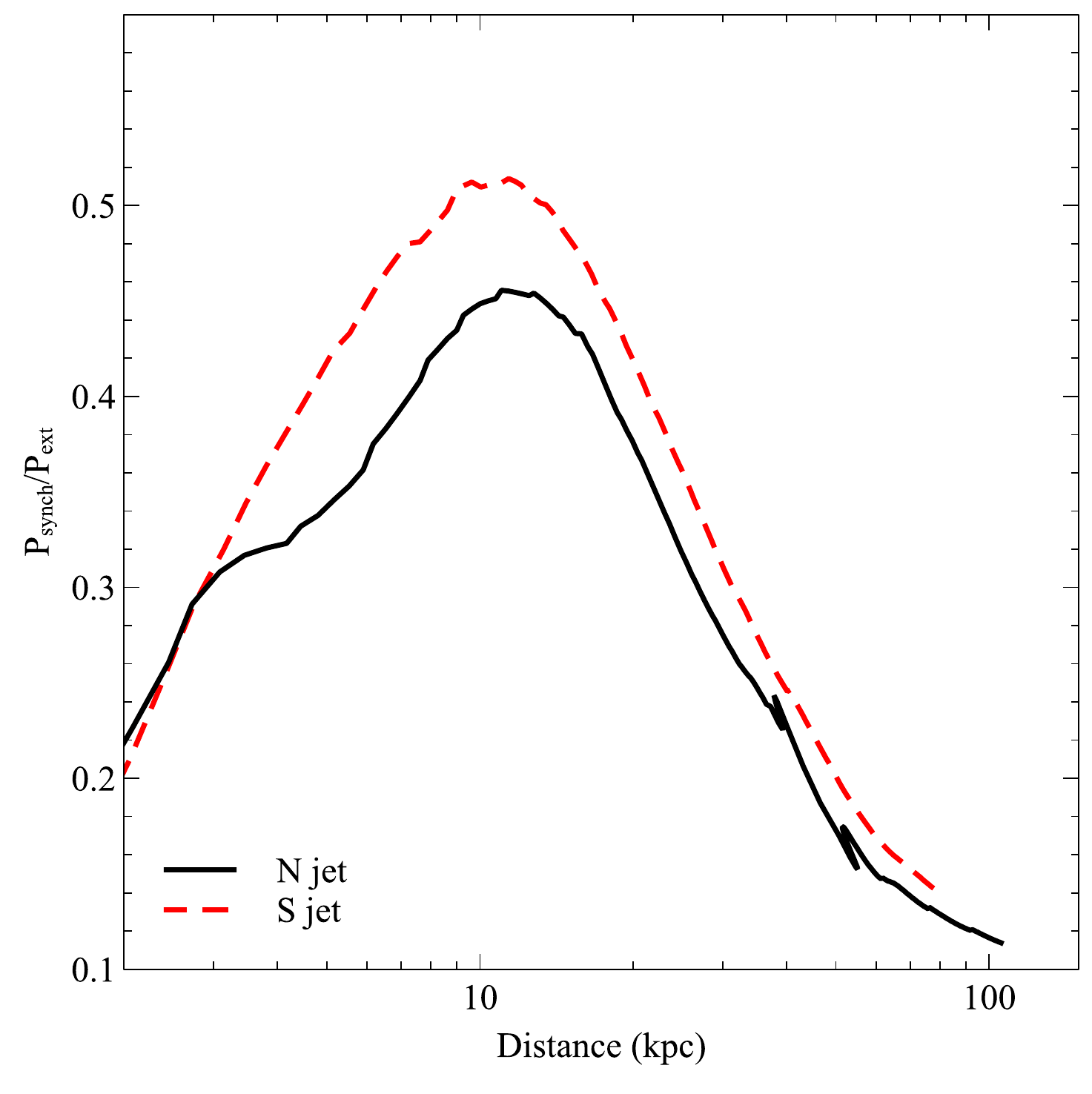}
%\hskip 1.0cm
 % \includegraphics[height=.25\textheight]{hyd_prat.eps}}}
  \caption{The fraction of required internal pressure that can be
    provided by the synchrotron-emitting components of the jets if at
    equipartition, as a function of distance from group/cluster
    centre for 3C\,31, showing that this component can provide a decreasing
    fraction of the jet pressure on scales of tens to hundreds of
    kpc. Line styles are as for Fig.~\ref{press1}.}
  \label{press}
\end{figure}

As previously stated, it is clear from Figs~\ref{press1} and
\ref{press} that in order for a departure from equipartition to be the
explanation for the apparent ``missing'' pressure in FRI lobes, the
jets must evolve further and further from the equipartition condition
as the source expands (apart from in the very inner parts -- we will
consider the implications of the differing behaviour in the inner
$\sim 10$ kpc of 3C\,31 in Section~\ref{inner})

\begin{figure}
\begin{center}
  \includegraphics[height=.3\textheight]{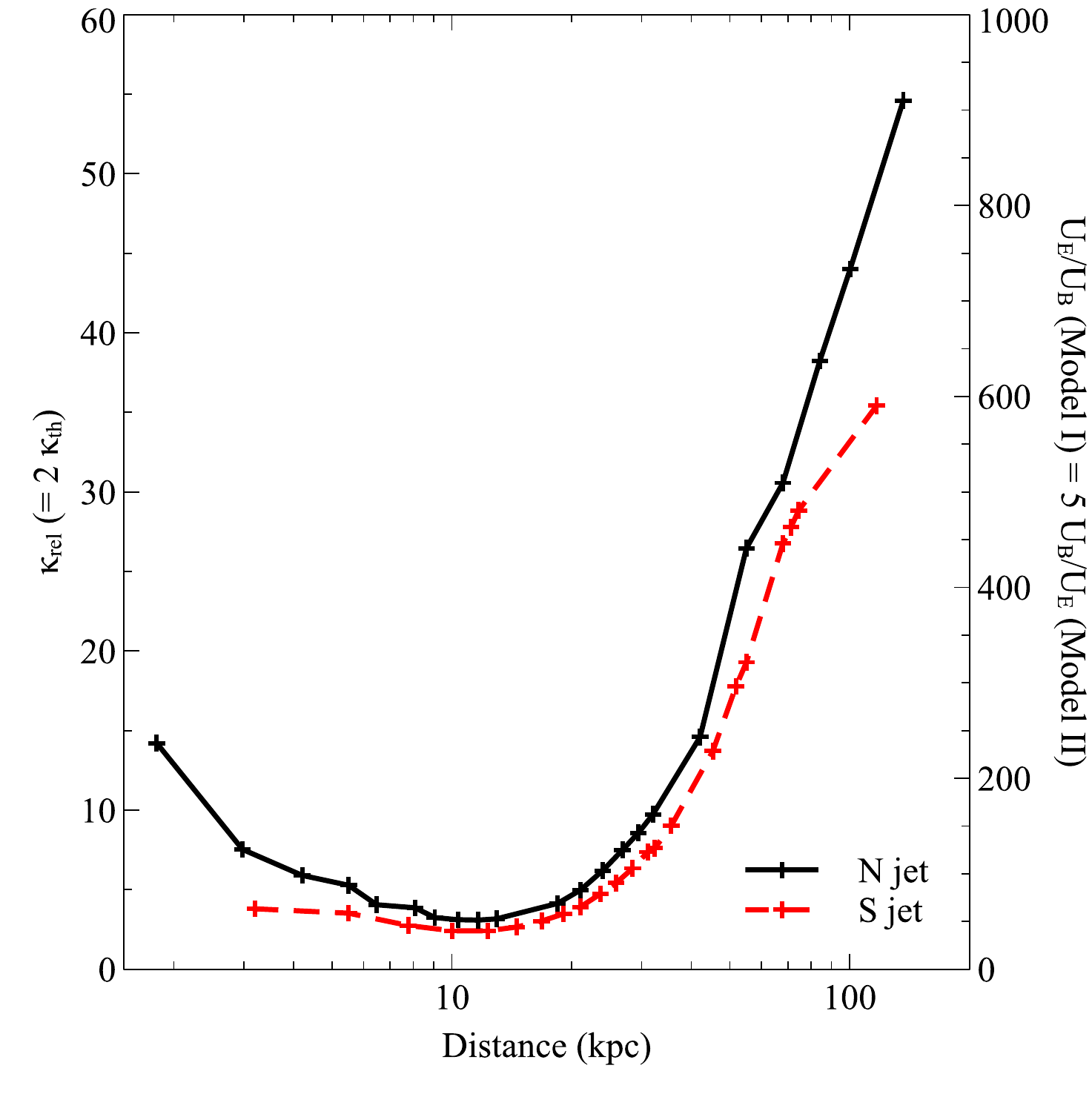}
  \caption{The evolution of the energetically dominant component of
    the 3C\,31 jets with distance, showing the ratio of lepton to
    magnetic field energy density for Model I, the inverse ratio for
    Model II, and the proton/ion content $\kappa$ for Models III and
    IV. Line styles are as for Fig.~\ref{press1}.}
  \label{protons}
\end{center}
\end{figure}

The energy densities and magnetic field strengths required in order
that the total energy density in the synchrotron-emitting plasma
should match the measured external pressure were determined by modelling
the electron energy distribution using the parameters discussed in
Section~\ref{radio}. Fig.~\ref{protons} shows the evolution of the
required energy ratio between magnetic fields and leptons required to
maintain pressure balance with the surrounding hot gas for Models I
and II.

In Model I (lepton domination) the particle energy dominates by a
factor $\sim 100$ in the inner regions, then, after an initial
decrease, increases to $\sim 500$ at hundred-kpc distances. For the
large electron densities required in this model, the predicted level
of X-ray inverse-Compton radiation from the radio jets and lobes would
be significant, and can be ruled out in a number of individual cases
\citep[e.g.][]{croston03b,hardcastle10b}.  In particular,
\citet{hardcastle10b} have examined in detail the constraints on
inverse Compton emission from Hydra A, and conclude that relativistic
electrons (and positrons) can contribute at most $\sim 6$ per cent of the internal
pressure of the radio lobes. We can therefore conclusively rule out
this explanation. For 3C\,31, we considered the outermost region of our
profile, and calculated the predicted level of X-ray inverse Compton
emission at 1 keV using the {\sc synch} code of \citet{hardcastle98c}
under the assumptions of Model I. We find that the observed residual
level of X-ray flux in this region after background subtraction is a
factor $\sim 2000$ times lower than the prediction of this model,
consistent with results for other FRI sources.

\begin{figure}
  \includegraphics[height=.3\textheight]{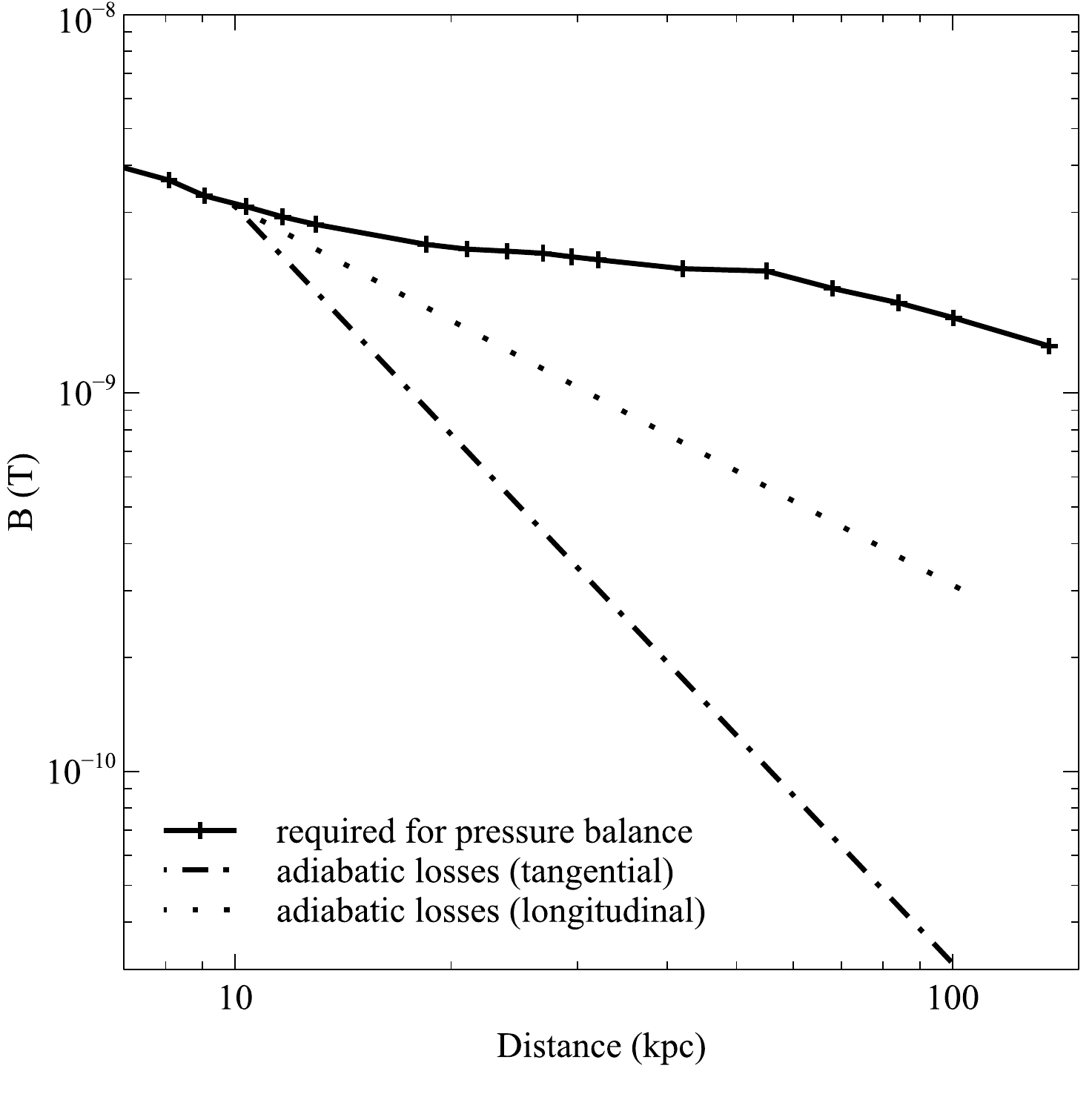}
  \caption{The magnetic field strength required as a function of
    distance in the case where magnetic field energy dominates the
    internal pressure (shown for the northern jet of 3C\,31). The dotted and dashed lines show the
    expected evolution of magnetic field strength due to adiabatic
    expansion for the case of a predominantly tangential and
    predominantly longitudinal field structure, respectively. Line styles are as for Fig.~\ref{press1}.}
  \label{bdom}
\end{figure}

In Model II (magnetic field domination), the energy ratio
$U_{B}/U_{E}$ evolves similarly to Model I, with the factor by which
the magnetic field dominates increasing from around 30 to $\sim 100$
by hundred kpc scale distances. Fig.~\ref{bdom} shows the magnetic
field strengths required as a function of distance to achieve pressure
balance in this model. The magnetic field strengths required are high
($\sim 10 - 40 \mu$G), decreasing by a factor of a few from the inner
parts to hundred kpc scale distances.  This model requires the
generation of magnetic field energy density along the source. The
dashed and dotted lines in Fig.~\ref{bdom} show the expected evolution
of magnetic field strength due to adiabatic expansion for the case of
a predominantly radial/toroidal and predominantly longitudinal field
structure, respectively \citep[e.g.][]{baum97}. A constant velocity
profile was assumed, which is conservative, as a decreasing velocity
would steepen the losses for the perpendicular components of
$B$. Hence a passively evolving magnetic field component is
inconsistent with the observations. The results shown in
Fig.~\ref{bdom} are not consistent with previously proposed models for
cylindrical jets with helical $B$ fields \citep[e.g.][]{nakamura06},
but such models are also inconsistent with FRI jet geometries and
  polarization structures \citep[e.g.][]{laing81,laing08a}.  The
requirement for a slow decrease in $B$ along the jets (despite lateral
expansion of the jet) could be consistent with a model in which
turbulence increasingly amplifies the magnetic field on large scales;
however, this would need to take place with no appreciable particle
acceleration for consistency with the radio constraints, and turbulent
amplification of magnetic fields beyond equipartition values is
challenging \citep{deyoung80}. Our results show that energy would have
to be being transferred from the particle population to the magnetic
field to a greater and greater extent at larger distances. This model
cannot be ruled out directly, but from the constraints on the model
given above we conclude that magnetic domination of the jets and lobes
is highly unlikely.

\subsection{Contributions from non-radiating particles (Models III and IV)}
\label{sec:protons}
The question of whether or not the inner jets of radio galaxies
consist of an electron-positron or electron-proton plasma is a
long-standing one, which has not yet been resolved satisfactorily,
despite substantial efforts over the past couple of decades
\citep[e.g.][]{ghisellini92,celotti93,wardle98,homan09}. On kpc scales, there is an
obvious additional source of non-radiating particles in the form of
material entrained into the jets from the surroundings: there is
substantial evidence for entrainment in FRI jets, and the standard
picture of FRI dynamics relies on entrainment to decelerate the jets
from relativistic to transonic speeds on scales of a few kpc
\citep[e.g.][]{bicknell94,laing02b}. In this section we consider models in
which relativistic protons (and/or ions (Model III) or thermal gas entrained from the surroundings (Model IV) dominate
the internal pressure.

The contribution from heavy particles (protons/ions) required to achieve pressure balance can be
determined straightforwardly under the assumption of equipartition of
energy between all particles (radiating and non-radiating) and
magnetic field. Details of this calculation for the cases of
relativistic and thermal gas are provided in
Appendix~A. In Fig.~\ref{protons} we plot the required ratio
of energy density in non-radiating particles to radiating particles for these two models.

\begin{figure}
  \includegraphics[height=.3\textheight]{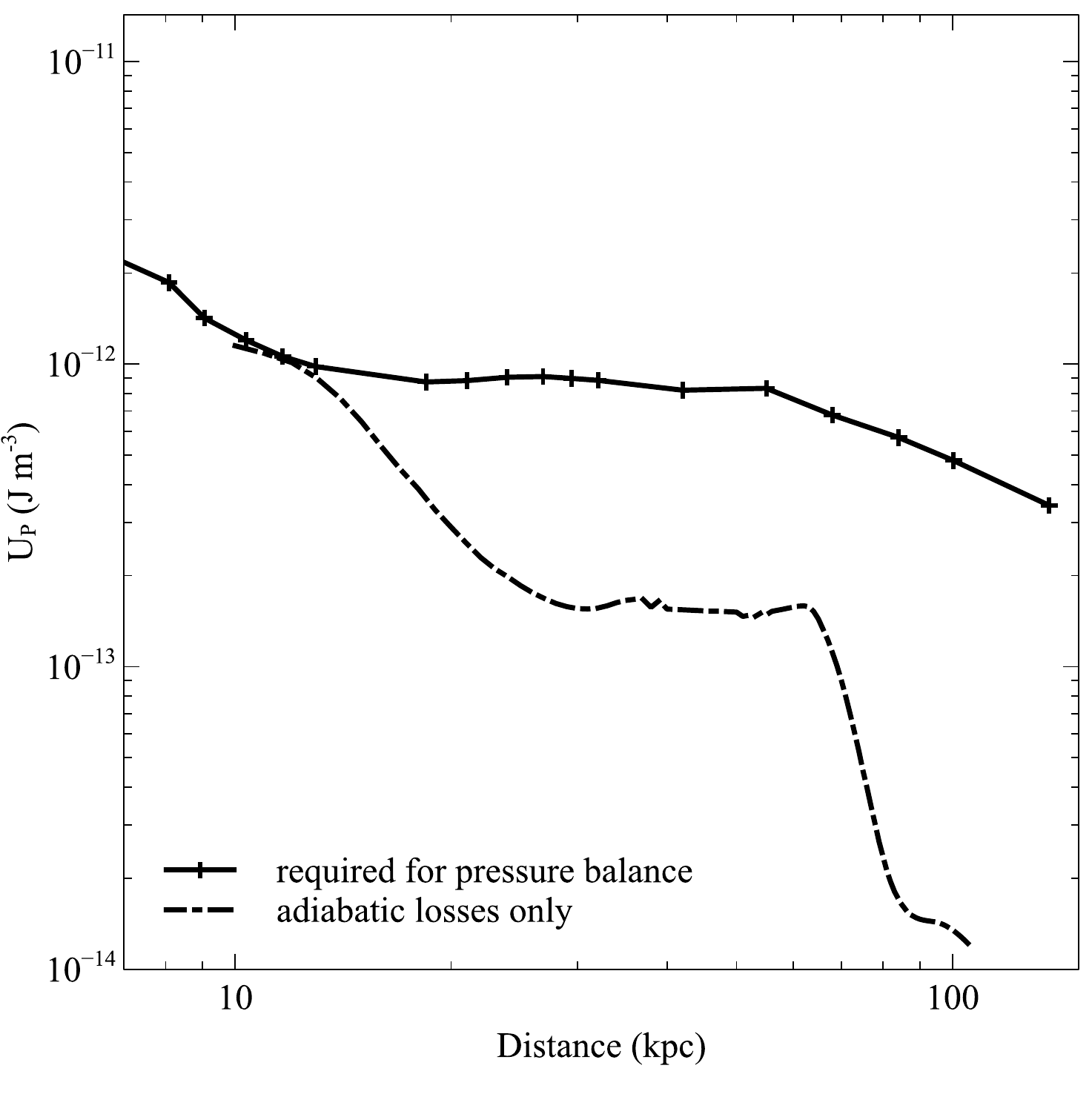}
  \caption{The energy density in relativistic protons and/or ions required to balance the
    external pressure (filled squares), shown for the northern jet of 3C\,31. The dashed and dotted lines indicate
    the expected evolution of energy density with distance along the
    source assuming adiabatic losses. Note that the flattening of the adiabatic model between 20 and 70 kpc is caused by the jet's cylindrical geometry in that region (see also Fig.~\ref{modelres})}
  \label{pdom}
\end{figure}

Fig.~\ref{pdom} shows the run of energy density in relativistic
protons (or ions) required to balance the external pressure for Model III,
assuming equipartition of particles (both radiating and non-radiating) with
magnetic field. If the electron population suffers significant
radiative losses (which do not affect the proton/ion energy density), it
might be expected that the relative energy density in protons (and/or ions) would
increase with distance, as required by the external pressure
data. However, if the energy is carried by relativistic protons
injected in the jets' inner regions, then their energy density would
be expected to evolve adiabatically with distance, in the absence of
significant radiative losses or particle acceleration. Fig.~\ref{pdom}
shows that the simplest version of Model III in which protons are
injected only in the inner jet is not viable, because the proton
energy density in this model decreases much less steeply with distance
than expected as a result of adiabatic losses (calculated from 10 kpc
outwards). We can therefore rule out a model in which protons injected
in the inner regions evolve passively along the jet. For relativistic
protons and/or ions to dominate the jets and lobes over their entire length,
significant particle acceleration is required on scales of tens to
hundreds of kpc (which must not significantly affect the lepton population).

A model in which entrainment of surrounding material leads to an
increasing thermal gas content as the jets evolve (whether or not they
initially contain relativistic protons), such as Model IV, is more
consistent with the data as it provides a simple explanation for the
decreasing energetic importance of the radiating particles as the jet
evolves. Fig.~\ref{protons} shows how the ratio of energy density in
non-radiating particles to radiating particles must evolve along the jet in this model. This
evolution of energy density could occur either by increasing
entrainment (via an increasingly large boundary layer), or by
increased heating/acceleration of entrained thermal gas. The required entrainment rate for Model IV can be
obtained by consideration of mass, momentum and energy flux conservation along
the jet. In the following section we develop a toy model to
investigate this scenario.

\subsection{An entrainment model on for 3C\,31}
\label{sec:entrain}
We model the region of jet between 12 kpc and 140 kpc, which is where
the X-ray constraints are tight while the uncertainties on jet
geometry are acceptable (beyond this distance further jet bends and
flaring making it difficult to constrain the geometry). The inner
boundary is chosen to be beyond the initial deceleration region
according to the model of \citet{laing02b}, so that relativistic
effects can be neglected. We assume Model IV, above, i.e. the
following assumptions hold: (1) the jet internal pressure, $P_{int}$,
balances the external pressure ($P_{ext}$, as measured from the X-ray
observations) at each radius; (2) the internal pressure has
contributions from magnetic field ($P_{B}$), synchrotron radiating
leptons ($P_{E}$), and thermal gas entrained from the environment
($P_{th}$); and (3) the magnetic field strength and energy density are
assumed to be in equipartition with the total particle energy density
(from synchrotron-radiating and non-radiating particles). We later discuss
the effects of relaxing the final assumption.

\begin{figure*}
  % \centerline{\hbox{
  \includegraphics[height=0.8\textheight]{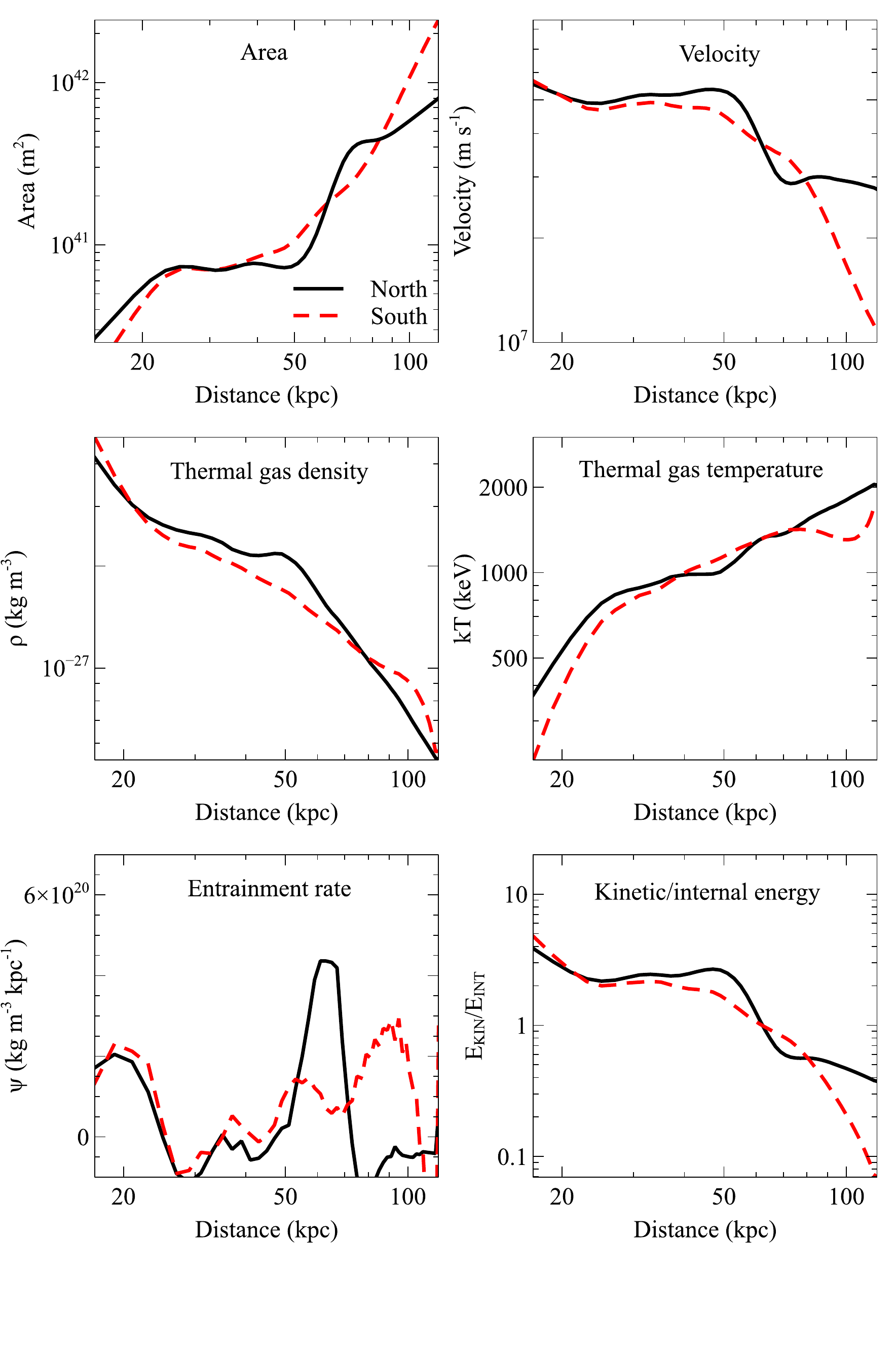}
\vskip -1.0cm
  \caption{Jet properties vs. distance for our entrainment model,
    assuming an initial temperature for the thermal component of 100
    keV. Top row:
    cross-sectional area (l) and velocity (r), middle row: gas density
    (l) and temperature (r) for the thermal component, bottom row:
    mass entrained per unit length (l) and ratio of kinetic to jet
    internal energy flux (r), all shown for the northern (black solid)
    and southern (red dashed) jets of
    3C\,31.}
  \label{modelres}
\end{figure*}

\begin{figure*}
  % \centerline{\hbox{
  \includegraphics[height=0.3\textheight]{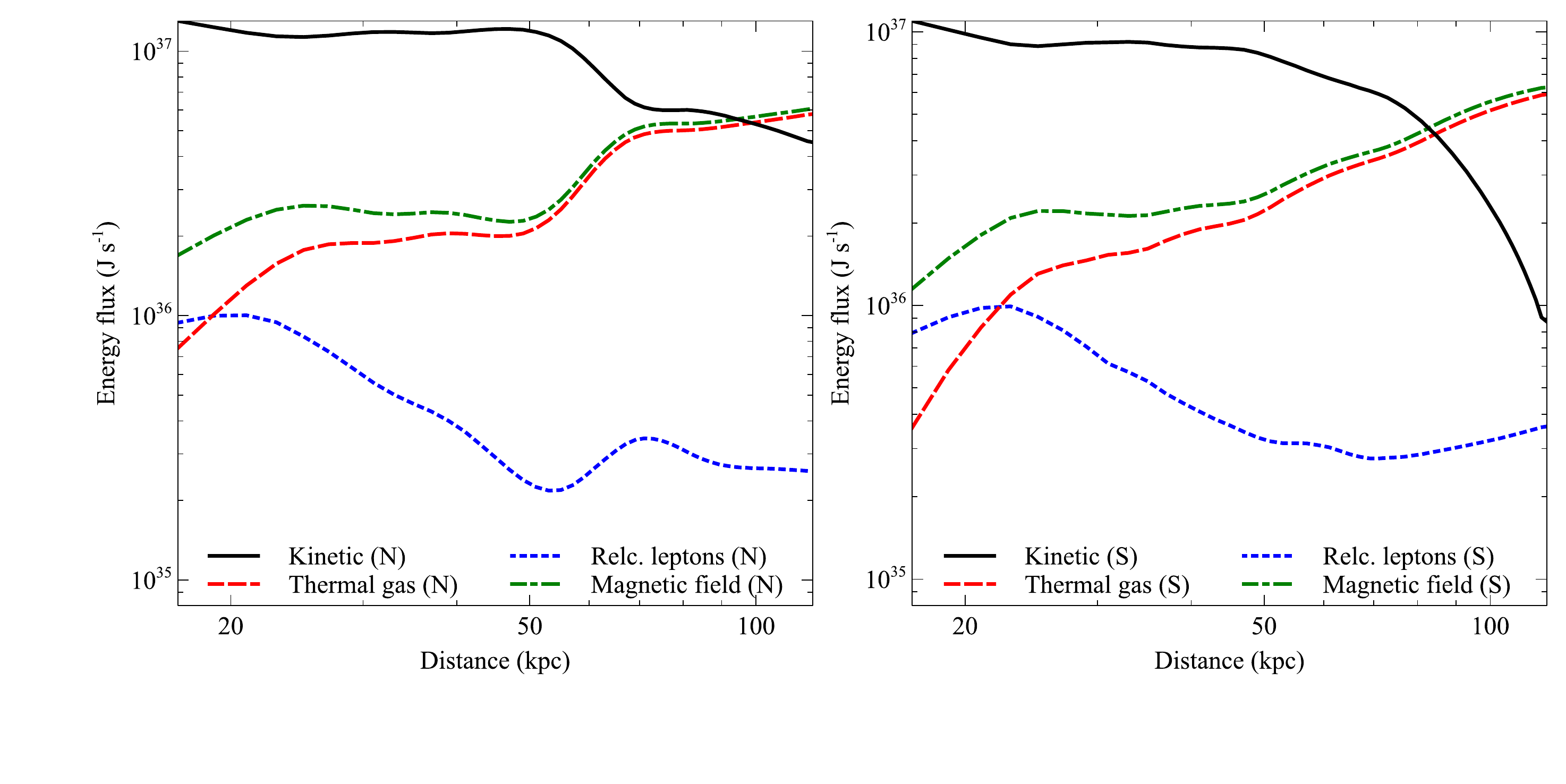}
  \caption{The evolution of energy flux with distance along the jet,
    for models with matched jet power, showing kinetic energy (black),
    magnetic field energy (green), internal energy of thermal
    particles (red) and of relativistic leptons (blue), with left and
    right panels indicating the northern and southern jets, respectively.}
  \label{energydiv}
\end{figure*}

By making use of the (non-relativistic) equations for conservation of
momentum and energy flux along the jet, the density and temperature of
the `missing' thermal component of the jet can be obtained, as
described in detail in Appendix~B. We require initial
conditions of density and velocity at the inner boundary. We take the
jet velocity of 3C\,31 at 12 kpc from the model of \citet{laing02b} as
our inner boundary condition, and assume a range of initial gas
temperatures. The choice of temperature for the thermal component at
12 kpc sets a boundary condition on the gas density (via the pressure
constraints), and hence determines the jet power. As discussed later,
we can therefore use the jet power as a consistency check on the
most appropriate choice of initial temperature.

Fig.~\ref{modelres} shows some illustrative results, with initial
conditions chosen to obtain jet powers matched for the two jets, and
in broad agreement with the model of \citet{laing02b} (this requires
initial temperatures at 12 kpc of 100 keV and 230 keV for the northern
and southern jets, respectively. In this model the behaviour of the
two jets is broadly similar, but with some differences driven by
variation in how the jet geometry evolves. The northern jet can be divided into
several regions on scales of tens to hundreds of kpc in which its
geometry differs. As shown in the top left panel of the figure, the
cross-sectional area initially increases steeply with distance, the
jet then becomes cylindrical between around 20 kpc to 60 kpc; and then
the jet radius increases again to 100 kpc scales and beyond. These
geometrical features show an interesting correspondence with bends in
the jet (occuring at both of those transition points), and with the
external pressure gradient, as the pressure profile flattens at around
20 kpc (plausibly moving from a galaxy-scale halo to a group-scale
atmosphere) and then steepens again between 50 and 100 kpc. The
density profile that results from an assumption of constant
temperature along the jet shows features that correspond to this
geometry, with an inner region of increasing density, followed by a
region of constant density and then a decreasing density in the outer
region as the jet/plume widens. Finally the bottom left panel shows
that in this model the entrainment must be fairly localized, with
large amounts of material ingested at the two transition points of
$\sim 20$ and $60$ kpc (note that these are distances along the jet
centre-line, rather than radial distances in the group atmosphere). At
other times the entrainment rate is low. The southern jet expands more
smoothly, and somewhat faster, consequently requiring entrainment to
be spread out over larger distances. At large distances the
cross-sectional area expands significantly more steeply than for the
northern jet, which leads to higher entrainment, deceleration, and
thermalization of kinetic energy.

The conservation-law analysis of \citet{laing02b} leads to an
entrainment rate at 12 kpc of $\sim 10^{20}$ kg s$^{-1}$
kpc$^{-1}$. The models shown in Fig~\ref{modelres} are consistent with
this level of entrainment; however, it is also possible that the
entrainment in our model results from fairly localised disruption of
the jet at its bends, which may be unconnected to the steadier
entrainment implied by the model of \citet{laing02b}, in which case
consistency with their measurement of entrainment rate is not required.

In our model the energy flux is primarily in the form of kinetic
energy in the inner parts of the jet, but is increasingly converted
into internal energy of the thermal (and presumably relativistic)
particles, as shown in Fig~\ref{energydiv}. The temperatures required
by our model, for realistic jet powers, are much higher than the
temperature of the surrounding gas, indicated that the entrained
material must be heated fairly rapidly by tapping the jet's kinetic
energy. We are assuming that all of the thermal material at a
particular distance in the jet has a single temperature, which is
simplistic; however, at any given position the majority of material
will have been in the jet for some time with recently entrained gas
comprising only a small fraction. We note that temperatures of $>100$
keV for entrained gas are consistent with the limits on the presence
of thermal material in cluster cavities obtained from limits on the
thermal X-ray emission due to this gas
\citep[e.g.][]{blanton03,sanders07}. The `thermal' component, although
very hot, remains (predominantly) sub-relativistic in this model,
although a non-thermal, relativistic tail cannot be ruled out.

Hence we conclude that our simple entrainment model is qualitatively
consistent with providing the dominant energetic contribution to the
jets and plumes of 3C\,31 on scales from 10 to 100 kpc. Most
interestingly, if entrainment does drive the source energetics, then much of the mass ingestion appears to be localised, and coincide
with regions where the jets bend and/or spread. In particular, the two
regions of the northern jet where entrainment takes place in our model
coincide with the flattening and
steepening of the external pressure profile, it is clear that the gas
distribution in the group environment determines the energetic
evolution of the radio-galaxy plasma on these scales.

Our assumption of equipartition of energy density between particles
and magnetic field may not be correct. We argue in Section~\ref{sec:depeq}
that non-equipartition models with no protons are unlikely to be
correct, but a model where thermal and relativistic particles together
dominate the energetics, with a lower magnetic field energy density,
cannot be ruled out. However, such a model would not strongly differ
in the qualitative picture for the evolution of thermal content of the
jet -- the radio synchrotron constraints mean that if the magnetic
field strength contributes a lower fraction of the internal energy
flux then the electron contribution must increase. Significant
entrainment would still be required at the locations seen in
Fig~\ref{modelres}, but the quantities of mass entrained and the
required temperature profile could be somewhat different.

\subsection{Geometrical uncertainties}
\label{geometry}
Uncertainty in the geometry of the radio jets, and in particular how
the jet orientation changes relative to the plane of the sky at the
observed jet bends, is potentially a major limitation of our
analysis. As discussed in Section~\ref{imp}, our main observational
result -- that the synchrotron-emitting components of the jet
contribute a decreasing fraction of the jet pressure, if at
equipartition -- is not affected by uncertainties in projection. In a
geometry with high inclination, the resulting larger synchrotron
emitting volume and lower external pressure acting on a particular
region due to larger radial distance in the cluster counteract each
other, which means that the overall result is largely unaffected. The
evolution of the non-radiating particle energy fraction in Models III
and IV (or of $U_{E}$ and $U_{B}$ in Models I and II) are therefore
qualitatively similar in any plausible geometry, even though the
numerical values will change somewhat. We do not attempt here to
derive precise constraints on the jet energy content at a particular
radius, but rather to develop a robust qualitative understanding of
how the components of the jet plasma evolve. Therefore, while we
acknowledge that the geometry is poorly constrained, our general
conclusions are robust.

\subsection{Uncertainties due to assumption electron energy distribution}

A further uncertainty comes from our lack of knowledge of the
low-energy electron distribution as a function of distance along the
jets. This will soon be remedied by ongoing work with LOFAR (Heesen et
al., in prep); however, at present we can only extrapolate to the
lowest frequencies from the radio spectrum at 330 MHz. As discussed in
Section~\ref{radio} we assumed a low-frequency spectral index of
$\alpha=0.55$ \citep[e.g.][]{laing13} and a value of $\gamma_{min} =
10$. Evidence for $\gamma_{min} \gg 1$ comes from the broad-band
spectra of hotspots \citep[e.g.][]{meisenheimer97,carilli99}; however, the
situation in FRI jets remains unknown. For the electron distribution
assumptions to significantly alter our results we would require an
evolution in the low-frequency properties of the jet with distance
from the nucleus. If $\gamma_{min}$ is determined by the particle
acceleration process that occurs in the inner jet, then it is
plausible that it could evolve to lower energies (e.g. via adiabatic
losses) at the plasma is advected downstream.  Alternatively, the
low-frequency spectral index could evolve to become steeper at larger
distances, but there is no indication that this is the case in the
existing 330-MHz data (e.g. the spectral index between 330 MHz and 608
MHz is $\sim 0.58$ for the outermost region we consider in the
northern jet).

We investigated the electron energy distribution that would be
required to achieve pressure balance in the outermost region of the
3C\,31 northern plume, assuming equipartition (the non-equipartition
cases having been considered previously). Simply reducing
$\gamma_{min}$ to 1, while extrapolating from the observed spectral
index of 0.55, is inadequate to achieve pressure balance. It would
necessary for the radio spectrum to steepen significantly below 330
MHz, to $\alpha > 0.9$, {\it and} to have a low-energy cut-off of
$\gamma_{min} = 1$ in order for the synchrotron emitting components to
provide all of the pressure within the lobes at this distance. As the
radiation from such a component is currently unobservable with
existing radio data, this scenario is effectively indistinguishable
from Models III and IV, above; however, it is difficult to reconcile
with particle acceleration models, and would require a second
relativistic particle population that has previously been
undetected. Such a dominant lepton population with $\gamma < 1500$,
emitting below 330-MHz, cannot currently be ruled out by existing radio
or X-ray inverse Compton constraints. We also cannot at this stage
rule out more complex models in which the spectral index (and
$\gamma_{min}$) vary while the contribution from thermal gas also changes
with distance, but we look forward to being able to test such models
in the near future with LOFAR data.

\subsection{Evolution in the inner jet region}
\label{inner}
We have focused mainly on the region of the jet beyond 10 kpc, where
it is thought to be subrelativistic and evolving into the group gas
environment. As shown in Figs.~\ref{press1} and \ref{press}, the
evolution of the jet plasma appears to be different in the region
inside 10 kpc. We have made no attempt to correct for the
effects of relativistic beaming in calculating our radio emissivity
profile as our focus is on the outer regions, but the effect of
``de-beaming'' the synchrotron emissivity [assuming the velocity model of
\citet{laing02b}] is a small decrease in the pressure of the
synchrotron components of the northern jet, and an increase in their
contribution for the southern jet. Hence this does not qualitatively
alter the behaviour of the northern jet, though it brings the southern
jet to have a roughly constant ratio of $P_{ext} / P_{synch}$ in the
inner region. 

If Model IV above is the correct explanation for the evolution of the
jet plasma on scales beyond 10 kpc, then other effects must be more
important in the inner region. One possibility is that the jet is
initially significantly electron (or relativistic electron and proton)
dominated (e.g. due to substantial particle acceleration in the inner
jet) before evolving towards equipartition between particles and
magnetic field, with entrainment taking over as an important mechanism
affecting the overall energetics from around 10 kpc. Such a model is
somewhat speculative, however, with the microphysics of energy
transfer between jet components poorly understood and difficult to
test.

\section{Conclusions}

We have shown that X-ray and radio measurements of external pressure
and internal pressure from radiating material as a function of
distance along the source can be used to distinguish between models
for the contents of radio lobes. Considering in detail the cases of
3C\,31 and Hydra A, we have shown that:
\begin{itemize}
\item The fractional contribution to the total energy budget from
  synchrotron-emitting components (relativistic leptons and magnetic
  field), if at equipartition, must decrease with distance from the
  central AGN.
\item A model in which the energetics are dominated by
  relativistic leptons can be ruled out by inverse-Compton limits.
\item Magnetic domination requires the magnetic field strength to
  remain close to constant along the jet, which is implausible given
  the jet geometry, due to the need to convert an increasing fraction
  of the jet energy into magnetic field as the jet evolves, without
  producing significant particle acceleration. 
\item A model in which relativistic protons/ions injected in the inner jet
  dominate the jet energetics and evolve adiabatically along the jet
  is ruled out.
\item Finally we have demonstrated that a simple entrainment model is
  consistent with the external pressure constraints and the evolution
  of radio emissivity, with regions of entrainment corresponding to
  locations of jet bending/disruption and changes in the external
  pressure profile. Such a model requires a high temperature for the
  entrained component, and an increasing temperature with distance,
  consistent with a rapidly decreasing kinetic energy flux of the jet
  being converted to particle and magnetic field internal energies.
\end{itemize}

The results presented here are based on consideration of a single
object, for which the highest quality radio and X-ray data on the
scales of interest are available. Our detailed pressure comparison for
Hydra\,A, as well as indications from less well constrained
comparisons for other objects \citep{evans05,hardcastle98e,worrall00}
and circumstantial evidence from observations of cluster cavities,
mean that it is plausible that our conclusion that an
entrainment-dominated model is favoured in 3C\,31 can be generalised
to low-power radio galaxies in general. In future work we will apply
these analysis methods to other systems with high-quality X-ray and
radio data, as well as incorporating new low-frequency radio
measurements to minimise uncertainties from extrapolation of the
electron energy distribution.

\section*{Acknowledgments}

JHC acknowledges support from the South-East Physics Network (SEPNet)
and from the Science and Technology Facilities Council (STFC) under
grant ST/J001600/1. We would like to thank Robert Laing for providing the
1.4-GHz map of 3C\,31. We would also like to thank the referee, Geoff
Bicknell, for a helpful report, which has enabled us to improve the paper.

\bibliographystyle{mn2e}
\bibliography{jhc}

%\singlecolumn
\section*{Appendix A. Method for calculating $\kappa$ (ratio of non-radiating particles
  to relativistic lepton energy density)}
\label{calc}
We assume:
\begin{itemize}
\item The jet internal pressure, $P_{int}$, balances the external
  pressure at each distance, $P_{ext}$, which is measured from the
  X-ray observations.
\item The internal pressure has contributions from magnetic field,
  $P_{B}$, synchrotron-radiating relativistic particles, $P_{E}$, and
  thermal gas entrained from the environment, $P_{th}$.
\item Pressure balance along the jet is described by the following
  relation between the external pressure and the internal energy
  densities of magnetic field, relativistic and non-relativistic
  (thermal) particles:
\begin{equation}
\label{pbalance}
P_{ext} = \frac{1}{3} U_{E} + \frac{1}{3} U_{B} + f U_{P}
\end{equation}
where $f$ is $1/3$ for relativistic protons/ions and $f = 2/3$ for thermal gas.
\end{itemize}

 The ratio of energy densities in non-radiating particles and
 synchrotron-emitting particles (relativistic leptons) is $\kappa$,
 i.e. $U_{P} = \kappa U_{E}$, leading to the following equations for
 the relativistic case:
\begin{equation}
\label{pbalance2}
3 P_{ext} = (1 + \kappa_{rel}) U_{E} + U_{B}
\end{equation}
and the thermal case:
\begin{equation}
\label{pbalance3}
3 P_{ext} = (2 \kappa_{th} + 1)U_{E} + U_{B}
\end{equation}

If we assume that the distribution of electron energy density is
described by a power law with index $p \neq 2$ (i.e. $N(E) =
N_{0}E^{-p}$), then the electron energy density is given by:
\begin{equation}
\label{urel}
U_{E} = \int_{E_{min}}^{E_{max}}{E N(E){\rm d}E} =
N_{0}\frac{E_{max}^{2-p} - E_{min}^{2-p}}{2 - p}
\end{equation}
where $N_{0}$ is the electron energy density normalization, $p$ is the
electron energy index, $E_{min}$ and $E_{max}$ are the lower and upper
cut-offs of the electron energy distribution.

We assume equipartition between magnetic field and all particles
(relativistic and non-relativistic), i.e. $U_{B} = (1 + \kappa)
U_{E}$, which leads to the standard expression for the equipartition magnetic field strength:
\begin{equation}
\label{beq}
B_{eq} = \left[\frac{2 \mu_{0} (1 + \kappa)J(\nu)\nu^{\frac{(p-1)}{2}}
  }{c_{1} (2 - p)}\left(E_{max}^{2-p} -
  E_{min}^{2-p}\right)\right]^{\frac{2}{p + 5}}
\end{equation}
where $J(\nu)$ is the synchrotron emissivity at a frequency $\nu$,
given by
\begin{equation}
\label{synch}
J(\nu) = c_{1} N_{0} \nu^{-\frac{(p-1)}{2}} B^{\frac{(p+1)}{2}}
\end{equation}
where $c_{1}$ is a constant \citep{longair94}:
\begin{equation}
\label{defc1}
c_{1} = k(p) \frac{e^{3}}{\epsilon_{0} c m_{e}} \left(\frac{m_{e}^{3}
  c^{4}}{e}\right)^{-\frac{(p-1)}{2}}
\end{equation}
where $k(p)$ is 0.050407 for $p=2$, 0.039484 for $p=2.2$, and 0.031547
for $p=2.4$. 
We can now make use of the pressure constraints derived earlier for
the relativistic proton/ion and thermal gas cases (Eqs,~\ref{pbalance2} and
\ref{pbalance3}, applying to Models III and IV, respectively) to get a
second expression for $B$. For Model III, substituting in for $U_{E}$ in Eq.~\ref{pbalance2}
gives:
\begin{equation}
\frac{B^{2}}{2 \mu_{0}} = 3 P_{ext} - \frac{(1 + \kappa) N_{0}}{2
  - p}\left[E_{max}^{2 - p} - E_{min}^{2 - p}\right]
\end{equation}
and substituting in Eq.~\ref{synch} gives:
\begin{equation}
\frac{B^{2}}{2 \mu_{0}} = 3 P_{ext} - \frac{(1 + \kappa) J(\nu)
  \nu^{\frac{(p - 1)}{2}} B^{-\frac{(p + 1)}{2}}}{c_{1} (2 -
  p)}\left[E_{max}^{2-p} - E_{min}^{2-p}\right]
\end{equation}
We can now substitute in our previously derived expression for the
equipartition $B$ field (Eq.~\ref{beq}):
\begin{multline}
\label{finaleq}
\left[\frac{2 \mu_{0} (1 + \kappa_{rel}) J(\nu) \nu^{\frac{(p-1)}{2}}}{c_{1}
    (2 - p)}\left(E_{max}^{2-p} -
  E_{min}^{2-p}\right)\right]^{\frac{4}{p + 5}} (2 \mu_{0})^{-1} =\\
3
P_{ext} - \frac{(1 + \kappa_{rel}) J(\nu)
  \nu^{\frac{(p - 1)}{2}}}{c_{1} (2 -
  p)}\left[E_{max}^{2-p} - E_{min}^{2-p}\right] \\
\left[\frac{2 \mu_{0} (1 + \kappa_{rel}) J(\nu) \nu^{\frac{(p-1)}{2}}}{c_{1}
    (2 - p)}\left(E_{max}^{2-p} -
  E_{min}^{2-p}\right)\right]^{-\frac{(p+1)}{(p + 5)}}
\end{multline}

This expression can be simplified to:
\begin{equation}
\kappa_{rel} = \left(\frac{3 P_{ext}}{2}\right)^{\frac{p+5}{4}} \left(2 \mu_{0}\right)^{\frac{p+1}{4}} c_{2}^{-1}
\end{equation}
where
\begin{equation}
c_{2} = \frac{J(\nu)\nu^{\frac{p - 1}{2}}}{c_{1} (2 - p)}\left[E_{max}^{2-p} - E_{min}^{-2-p}\right]
\end{equation}

For the thermal case (Model IV), a similar expression can be derived from Eq~\ref{pbalance3}, with a slightly difference dependence on $\kappa$:
\begin{multline}
\label{finaleq2}
\left[\frac{2 \mu_{0} (1 + \kappa_{th}) J(\nu) \nu^{\frac{(p-1)}{2}}}{c_{1}
    (2 - p)}\left(E_{max}^{2-p} -
  E_{min}^{2-p}\right)\right]^{\frac{4}{p + 5}} (2 \mu_{0})^{-1} =\\
3
P_{ext} - \frac{(1 + \kappa_{th}) J(\nu)
  \nu^{\frac{(p - 1)}{2}}}{c_{1} (2 -
  p)}\left[E_{max}^{2-p} - E_{min}^{2-p}\right] \\
\left[\frac{2 \mu_{0} (1 + \kappa_{th}) J(\nu) \nu^{\frac{(p-1)}{2}}}{c_{1}
    (2 - p)}\left(E_{max}^{2-p} -
  E_{min}^{2-p}\right)\right]^{-\frac{(p+1)}{(p + 5)}}
\end{multline}
which simplifies to:
\begin{multline}
3 P_{ext} = (2 \mu_{0})^{-\frac{p+1}{p+5}} c_{2}^{\frac{4}{p+5}}\\
\left[(1 + \kappa_{th})^{\frac{4}{p+5}} + (2 \kappa_{th} + 1)(1 + \kappa_{th})^{-\frac{p+1}{p+5}}\right]
\end{multline}
For both models we have now have an equation that contains only one
unknown, $\kappa$. For an observed $S(\nu^{\prime})$ and $P_{ext}$,
the value of $\kappa$ can therefore be obtained (numerically, in the
thermal case), which also allows $B$, $U_{E}$, and $U_{p}$, and
finally the thermal gas density in the jet for a given assumed
temperature, to be obtained.

\section*{Appendix B. Details of entrainment calculations}
\label{app:method}
For a steady-state jet, the evolution of the dynamics and energy
content of the jet can be described by the equations of conservation
of momentum flux and energy flux. We assume that the jet velocity is
non-relativistic, which is appropriate for the region of jet
considered in this analysis. We consider a region of jet between
distance $l_{1}$ and $l_{2}$ from the nucleus. The conservation of
momentum flux, $\Pi = \rho v^{2} A $, is described by:
\begin{equation}
\rho_{2} v_{2}^{2} A_{2} = \rho_{1} v_{1}^{2} A_{1} + \Pi_{\rm buoy}
\label{pcons}
\end{equation}
where $\rho_{1,2}$, $A_{1,2}$ and $v_{1,2}$ are the gas density,
cross-sectional area and velocity of the jet, respectively, and
$\Pi_{\rm buoy}$ is the change in momentum flux due to the buoyancy force
acting on the jet (see below).

The conservation of energy flux can be described by \citep[cf.][]{bicknell94}:
\begin{multline}
\left(\frac{1}{2} \rho_{2} v_{2}^{2} + U_{2} + P_{2}\right) v_{2}
A_{2}= \left(\frac{1}{2} \rho_{1} v_{1}^{2} + U_{1} + P_{1}\right) v_{1}
A_{1}
\label{econs}
\end{multline}
where $P_{1,2}$ is the total internal pressure (assumed to match the
external pressure at the given radius),  and the internal energy terms, $U_{1,2}$ are given by:
\begin{equation}
U_{i} = \frac{3}{2} P_{i} + U_{e} + U_{B},
\end{equation}
i.e. including terms for the internal energy carried by thermal
particles, relativistic leptons and magnetic field, respectively.

With suitable initial conditions, the run of external pressure and of
$\kappa$ determined from the analysis in Section~\ref{sec:protons},
Equations~\ref{pcons} and \ref{econs} can be solved for the unknowns
$\rho_{2}$ and $v_{2}$ (where the mean particle mass $\mu - 0.6$, as
appropriate for entrained ICM gas). The gas temperature of the thermal
material is also determined via $P_{therm,i} = (\rho_{i} / \mu m_{H})
k T_{i})$ where $P_{thermal} = (2/3) U_{thermal} = (2 \kappa_{i} / 3)
U_{E,i}$ is determined from the analysis in Section~\ref{sec:protons}.

As discussed in Section~\ref{sec:entrain} as initial conditions we
assume $v_{12kpc} = 6 \times 10^{7}$ m s$^{-1}$, and test a range of
initial temperature values, which together with $P_{12kpc}$ determine
$\rho_{1}$. Equation~\ref{pcons} is rearranged for $v_{2}$, and then substituted
into Equation~\ref{econs}. A standard root-finding algorithm can then
be used to solve for $\rho_{2}$. The temperature of the thermal
material in region 2 is then determined as explained above.

The buoyancy term $\Pi_{{\rm buoy}}$ in Equation~\ref{pcons} is determined as follows. The
buoyant force acting on the jet material between $l_{1}$ and $l_{2}$
is given by:
\begin{equation}
F_{buoy} = - \Delta m g 
\end{equation}
where $\Delta m$ is the mass of the surrounding material displaced by
the chunk of jet material between $l_{1}$ and $l_{2}$, i.e. $\Delta m
= \Delta m_{env} - \Delta m_{jet}$, and $g$ is the acceleration due to gravity:
\begin{equation}
g = \frac{G m(l)}{l^{2}}
\end{equation}
where $m(l)$ is the enclosed gas mass within the galaxy cluster at radius
$l$. In this case we can assume that $\Delta m_{env} >> \Delta m_{jet}$, and so we take $\Delta m = \Delta m_{env} = \rho_{env} (l) A(l) \delta l$, where $\rho_{env}(l)$ is the external gas density at distance $l$.

The change in momentum is given by:
\begin{equation}
\Delta p = F_{buoy} \Delta t
\end{equation}
where $\Delta t$ is the interval during which material travels from
$l_{1}$ to $l_{2}$. Hence $\Delta l = v(l) \Delta t$.

The change in momentum flux due to buoyancy is therefore:
\begin{equation}
\Pi_{buoy} = \Delta p v = \int_{l_{1}}^{l_{2}}{F_{buoy} {\rm d}l}
\end{equation}
Using the equation of hydrostatic equilibrium, $F_{buoy}$ can be
expressed in terms of the external pressure gradient:
\begin{equation}
F_{buoy} = -\frac{G m(l) \rho_{env} (l) A(l) \Delta l}{l^{2}} = \frac{{\rm
    d}P_{ext}}{{\rm d}l} A(l) \Delta l
\end{equation}
So
\begin{equation}
\Pi_{buoy} = \int_{l1}^{l1}{\frac{{\rm
    d}P_{ext}}{{\rm d}l} A(l) {\rm d}l} 
\end{equation}
Hence the buoyancy term in Equation~\ref{pcons} can be evaluated using our measured external pressure gradient (e.g. Fig~\ref{press1}) and jet geometry.

Finally, the mass entrainment rate can be determined from conservation of mass flux as follows:
\begin{equation}
\rho_{2} v_{2} A_{2} = \rho_{1} A_{1} v_{1} + \Psi_{2}
\end{equation}
where $\Psi_{2}$ is the mass entrained per unit time in the region between $l_{1}$ and $l_{2}$. 

\end{document}